\documentclass[twocolumn, tighten, times]{aastex631}
\usepackage{comment}
\shorttitle{Cosmogenic Origin of KM3-230213A}
\shortauthors{Boxi, Das \& Gupta}

\graphicspath{{./}{figures/}}

\usepackage[dvipsnames]{xcolor}

\usepackage{lipsum}
\usepackage{float}

\begin{document}


\title{Cosmogenic Origin of KM3-230213A: Delayed Gamma-Ray Emission from A Cosmic-Ray Transient}
\author[0009-0004-1305-9578]{Sovan Boxi}
\email{sovanboxi@rrimail.rri.res.in}
\affiliation{Raman Research Institute,
C.V. Raman Avenue, 5th Cross Road, Sadashivanagar, Bengaluru, Karnataka 560080, India}
\author[0000-0001-5796-225X]{Saikat Das}
\affiliation{Department of Physics, University of Florida, Gainesville, FL 32611, USA}
\email{saikatdas@ufl.edu}
\author[0000-0002-1188-7503]{Nayantara Gupta}
\affiliation{Raman Research Institute,
C.V. Raman Avenue, 5th Cross Road, Sadashivanagar, Bengaluru, Karnataka 560080, India}
\email{nayan@rri.res.in}

\begin{abstract}
The highest-energy cosmic neutrino detected by the ARCA detector of KM3NeT has reignited the quest to pinpoint the sources of ultrahigh-energy cosmic rays (UHECRs; $E\gtrsim 0.1$ EeV). By uncovering the associated multimessenger signals, we investigate the origin of the 220 PeV $\nu_\mu$ event KM3-230213A from an unknown transient that accelerated cosmic rays to $\sim 10$ EeV. Unlike an astrophysical origin, where the $\nu_\mu$ is produced inside the source, here we consider UHECR protons that escape the source interact with the cosmic background radiation, producing a PeV-EeV cosmogenic neutrino spectrum. The secondary $e^\pm$ and $\gamma$-rays initiate an electromagnetic cascade, resulting in a cosmogenic $\gamma$-ray spectrum. The latter peaks at a delayed time of $\gtrsim 10^4$ years compared to the light travel time from the transient to observer, due to deflection of charged particles in the extragalactic magnetic field (EGMF). Our results shed light on the nature of the UHECR source for the $\nu_\mu$ event and provide crucial insights into the detection of multi-TeV $\gamma$-rays of cosmogenic origin from similar past cosmological transients. Using the $\gamma$-ray sensitivity of currently operating and next-generation imaging atmospheric Cherenkov telescopes, the flux and time-delay distribution can constrain the source distance. We further show that the detection of such a $\gamma$-ray signal above the background depends on the EGMF strength. Together with the non-detection of coincident spatial or temporal photon counterparts at the current epoch, this detection is the first compelling candidate for a sub-EeV cosmogenic neutrino.

\end{abstract}


\section{Introduction} 

The origin of ultrahigh-energy cosmic-ray (UHECRs; $E\gtrsim 10^{17}$ eV) acceleration is a long-standing puzzle that has intrigued astronomers for several decades \citep[see, e.g.,][]{Anchordoqui:2018qom}. UHECR sources reveal themselves through secondary $\gamma$-rays and neutrinos produced in their interactions \citep[see, e.g.,][]{Stecker_1979, Kotera_2010, AlvesBatista:2019tlv}. While astrophysical neutrinos originate directly from the acceleration sites of cosmic rays, cosmogenic neutrinos are produced during their cosmological propagation from the source to Earth. In the latter case, due to UHECR interactions with cosmic background photons, neutrinos are accompanied by an associated flux of GeV-TeV cosmogenic $\gamma$-rays, irrespective of the optical depth at the source. The IceCube observatory has detected an astrophysical neutrino spectrum in the TeV-PeV energy range at $\gtrsim5\sigma$ statistical significance \citep{IceCube:2013low, IceCube:2014stg}. However, the characteristic spectrum of cosmogenic neutrinos may extend to much higher energies depending on the energy of the target photons for photohadronic interactions. 

On February 13, 2023, the ARCA detector of KM3NeT, operating for a total time of 287.4 days, recorded an exceptionally 
energetic track-like event, designated KM3-230213A \citep{KM3NeT:2025npi, KM3NeT:2025ccp}. The reconstructed muon energy is estimated to be $120^{+110}_{-60}$~PeV at 68\% C.L., implying a parent neutrino energy of $\sim 220$~PeV, assuming charged-current interaction of a muon neutrino. The production of a 220 PeV $\nu_\mu$ event requires a parent cosmic-ray energy of $\approx4.4$ EeV, indicating an extragalactic origin \citep{KM3NeT:2025aps}. Several candidate sources have been proposed for this event, including year-long transients \citep{Neronov:2025jfj}, blazars \citep{Dzhatdoev:2025sdi, Yuan:2025zwe}, jet-red giant interaction \citep{deClairfontaine:2025gei}, electroweak vacuum instability \citep{Sakharov:2025oev}, and cosmological scenarios such as superheavy dark-matter decay and sterile neutrinos \citep{Narita:2025udw, Kohri:2025bsn, Choi:2025hqt, Murase:2025uwv, Dev:2025czz, Aloisio:2025nts}. For an astrophysical origin, no statistically significant Fermi-LAT source is identified in the analysis of the electromagnetic cascade of $\gamma$-rays originating from the source \citep{Crnogorcevic:2025vou}. 

A cosmogenic neutrino hypothesis, assuming a proton contribution to the highest-energy UHECR spectrum and a strong redshift evolution of the source population, has been proposed \citep{KM3NeT:2025vut, Zhang:2025abk, Kuznetsov:2025ehl}. However, the Fermi-LAT diffuse isotropic $\gamma$-ray background (IGRB) severely constrains the cumulative contribution from such a population \citep{Cermenati:2025ogl}. A joint fit to neutrino data from IceCube and KM3NeT, along with the UHECR spectrum and composition data from the Pierre Auger Observatory \citep{PierreAuger:2023kgv}, indicates a distinct cosmic-ray source population \citep{Muzio:2025gbr}. The non-detection of a delayed photon signal from electromagnetic cascades of UHE $\gamma$ rays injected by the neutrino source can constrain the intergalactic magnetic field \citep{Fang:2025nzg}. For blazar cosmogenic scenarios, a source redshift $z \lesssim 1$ remains compatible with the typical Eddington luminosity of active galactic nuclei \citep{Das:2025vqd}. 

In this letter, we analyse the cosmogenic fluxes from a transient point source. Here, the unknown transient phenomenon may be a short-lived astrophysical event characterized by an episode of increased activity (and thus increased cosmic-ray luminosity) relative to the emission during the baseline quiescent state. The all-sky diffuse cosmogenic neutrino flux arises from integrating over all point sources at all redshifts. However, we assume that the transient source considered in our analysis belongs to a subdominant unresolved UHECR source population that injects protons, consistent with Fermi-LAT IGRB constraints on a proton-dominated UHECR source model. We present the prospects for detecting the associated cosmogenic $\gamma$-ray flux, peaking at a much later time, if the neutrino event originated from a UHECR interaction. We consider various source distances and examine the time-delay distributions of neutrinos and $\gamma$-rays. Our analysis constrains the source distance by the absence of coincident electromagnetic counterparts at the current epoch. This analysis also provides insight into orphan very-high-energy (VHE; $E_{\gamma}\gtrsim0.1$ TeV) $\gamma$-ray detection as a delayed signal from a past transient and hints at the nature of such sources, which produce a neutrino flux comparable to that of the KM3-230213A event at a much earlier time.

In Sec.~\ref{sec:simulation}, we summarize our model considerations and the simulation setup used to analyze cosmogenic neutrino and $\gamma$-ray fluxes. We present our results in Sec.~\ref{sec:results} and discuss the implications in Sec.~\ref{sec:discussions}. We draw our conclusions in Sec.~\ref{sec:conclusion}.

\section{Cosmogenic fluxes and Simulation Setup\label{sec:simulation}}

UHECRs beyond the Greisen-Zatsepin-Kuzmin (GZK) cutoff energy of $5\times10^{19}$ eV undergo resonant photopion interaction with the cosmic microwave background (CMB) photons \citep{Greisen_1966, Zatsepin_1966}, leading to a cosmogenic neutrino spectrum peaking at $\gtrsim 1$ EeV, while lower energy cosmic rays dominantly interact with the extragalactic background light (EBL) comprised of infrared/optical/ultraviolet photons. In our model, we consider that the source injects UHECR protons in the energy range 1--10 EeV with a power-law injection spectrum given by $dN/dE\sim E^{-2}$, sufficient to produce neutrinos in the energy uncertainty range of the KM3NeT event. 

We employed the Monte Carlo simulation framework \textsc{CRPropa3} \citep{AlvesBatista:2016vpy, AlvesBatista:2022vem} to simulate the 3-D propagation and interactions of UHECRs and to track the development of the resulting electromagnetic (EM) cascades initiated by secondary $e^\pm$ and $\gamma$-rays. We set up a random turbulent extragalactic magnetic field (EGMF) with a Kolmogorov power spectrum over length scales of 100 kpc to 5 Mpc, and a turbulence coherence length of $\approx1$ Mpc. We choose an optimistic value of the RMS field strength $B_{\rm RMS} \approx 10^{-5}$ nG. Fermi observations of TeV blazars constrain the EGMF to be as low as $\sim10^{-7}$ nG \citep{Neronov_2010}. The magnetic field is periodically repeated in space to cover an arbitrary simulation volume. Due to computational cost, we calculate our cascade photon flux down to 10 GeV. Lower-energy photons are expected to arrive at higher deflection angles and are difficult to associate with the transient that produces such a neutrino event. We calculate the time delay of the cosmogenic $\gamma$ ray fluxes with respect to the light travel time from the source to the observer. This is much longer than that due to $\gamma$-ray induced cascade resulting from an astrophysical origin, as shown in \cite{Fang:2025nzg}.

We include all relevant energy-loss processes, photopion production, Bethe–Heitler pair production in the CMB and EBL, adiabatic losses due to cosmic expansion, and $\beta$-decay of secondary neutrons. We use the \cite{Gilmore_2012} model for EBL. The secondary EM particles lose energy via pair production, including double and triplet pair production, inverse Compton scattering of cosmic background photons, and synchrotron radiation due to the EGMF. We also include the universal radio background from \cite{Nitu:2020vzn}, crucial for UHE photon interactions. 

\begin{figure}[t]
    \centering
    \includegraphics[width=0.45\textwidth]{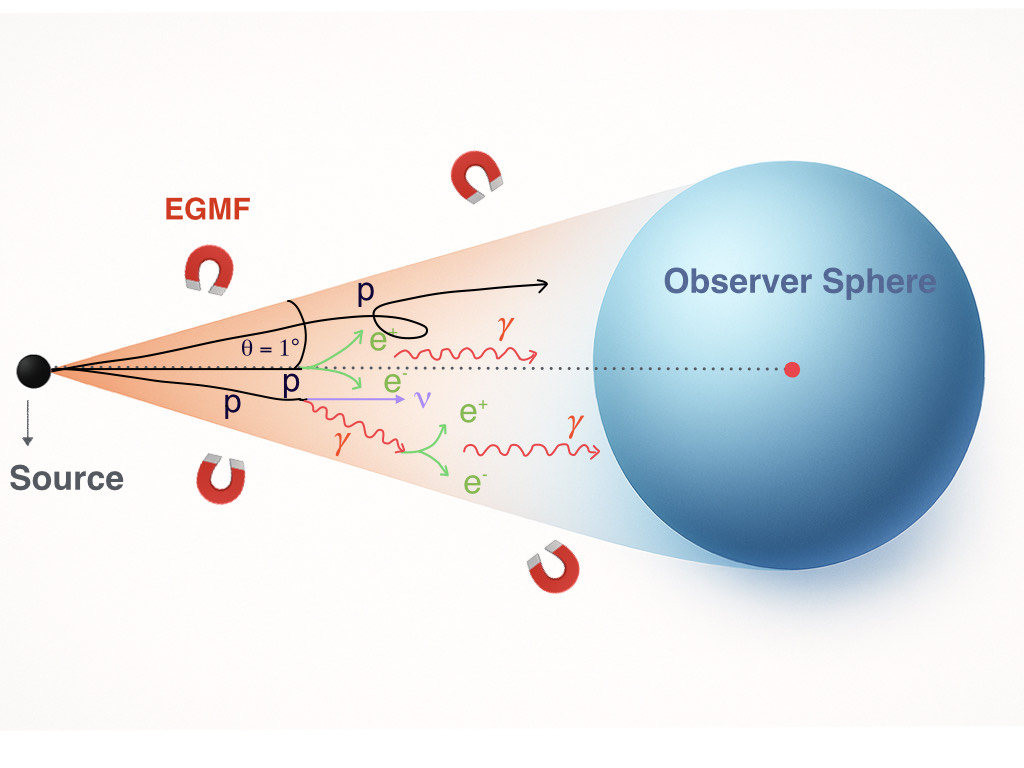}
    \caption{Schematic view showing source position and observer. We use an observer sphere to collect secondary events with a maximum deflection of $1^\circ$ in our 3-D simulations.
    }
    \label{fig:schematic}
\end{figure}

UHECRs and secondary $e^\pm$ are deflected in cosmic magnetic fields, causing a delay in the cosmogenic $\gamma$-ray signal. For neutrinos, UHECR interactions with the EBL are the dominant contribution in our chosen energy range. The corresponding cosmogenic neutrino flux is normalized to the flux predicted by the KM3NeT-only fit. The per-flavor diffuse flux is $E_\nu^2\Phi_{\nu+\overline{\nu}}^{\rm 1f}=5.8\times10^{-8}$ GeV cm$^{-2}$ s$^{-1}$ sr$^{-1}$ assuming a $E_\nu^{-2}$ spectrum in the central 90\% energy range of 72 PeV - 2.6 EeV \citep{KM3NeT:2025ccp}. Since the flux corresponds to a single observed event, we consider the KM3NeT-only fit for our analysis of a transient point source and convert it to point-like flux by multiplying by $4\pi$. The initial emission direction of UHECRs is collimated along our line of sight. Secondary neutrinos and $\gamma$-rays are collected on the surface of an observer sphere of radius such that a maximum deflection of $1^\circ$ is recorded, i.e., the observer sphere increases in size as the distance to the source increases \citep{Elyiv_2009}. The KM3NeT-ARCA event has a 99\% C.L. angular uncertainty of $3^\circ$. However, the Fermi-LAT resolution is $0.15^\circ$ for photons of energy $>\!30$ GeV. 

\section{Results \label{sec:results}}

Fig.~\ref{fig:schematic} shows a schematic representation of the scenario considered here. The straight line path from the source to the observer corresponds to the case of zero time delay between neutrino and $\gamma$-ray signals. UHE protons can be deflected by the EGMF, leading to arrival delays in observed signals \citep{Kotera:2008ae, Dermer:2008cy, Murase_2012}. Neutrinos, although propagating rectilinearly once produced, may still inherit delays due to proton propagation time before neutrino production and the transverse offset between neutrino production site and the source-observer line of sight. Cosmogenic photons experience additional delays due to the EM cascade, in which pair production and inverse-Compton scattering lengthen the effective path length through successive cascade generations.  

The additional delay in the cosmogenic photon spectrum depends on the deflection of secondary electrons during the inverse-Compton energy loss time of $e^\pm$ pair on CMB, $t_{\rm IC}=3m_ec/4\gamma_e\sigma_Tu_{\rm CMB}\approx 0.5\times10^{13}(\gamma_e/10^7)^{-1}$~s at $z=0.1$ \citep{Razzaque:2004cx}. The deflection angle of the pairs can be approximated as $\theta_B \approx ct_{\rm IC}/R_L \approx 8.8\times10^{-2}(\gamma_e/10^7)^{-2}(B_{\rm rms}/10^{-14} \text{\ G})$, where $R_L=\gamma_em_ec^2/eB_{\rm rms}$ is the Larmor radius of the pairs in the EGMF. For a pencil beam, this corresponds to $\Delta t_B=t_{\rm IC}\theta_B^2/2 \approx 600 \text{\ yrs} \times (\gamma_e/10^7)^{-5}(B_{\rm rms}/10^{-14} \text{\ G})^2$. However, in reality, the $\gamma\gamma$ pair production length plays a crucial role, and $e^\pm$ pairs have a finite spreading. Hence, the time delay could be written as $\Delta t =\lambda_{\rm \gamma\gamma}\theta_B^2/2c \approx 3.8\times 10^5$~yrs $\times(\gamma_e/10^7)^{-4}\times(n_{\rm CIB}/0.1 \text{\ cm}^{-3})^{-1}\times(B_{\rm rms}/10^{-14} \text{\ G})^2$ at $z\approx0.1$, where $n_{\rm CIB}$ and $\lambda_{\rm \gamma\gamma}$ are the infrared photon number density and the energy loss length due to pair-production \citep{Murase:2008pe}. Whereas neutrinos, once produced, travel along a rectilinear path undeflected by the magnetic field and unattenuated by cosmic background radiation.

\begin{figure}[t]
    \centering
    \includegraphics[width=0.45\textwidth]{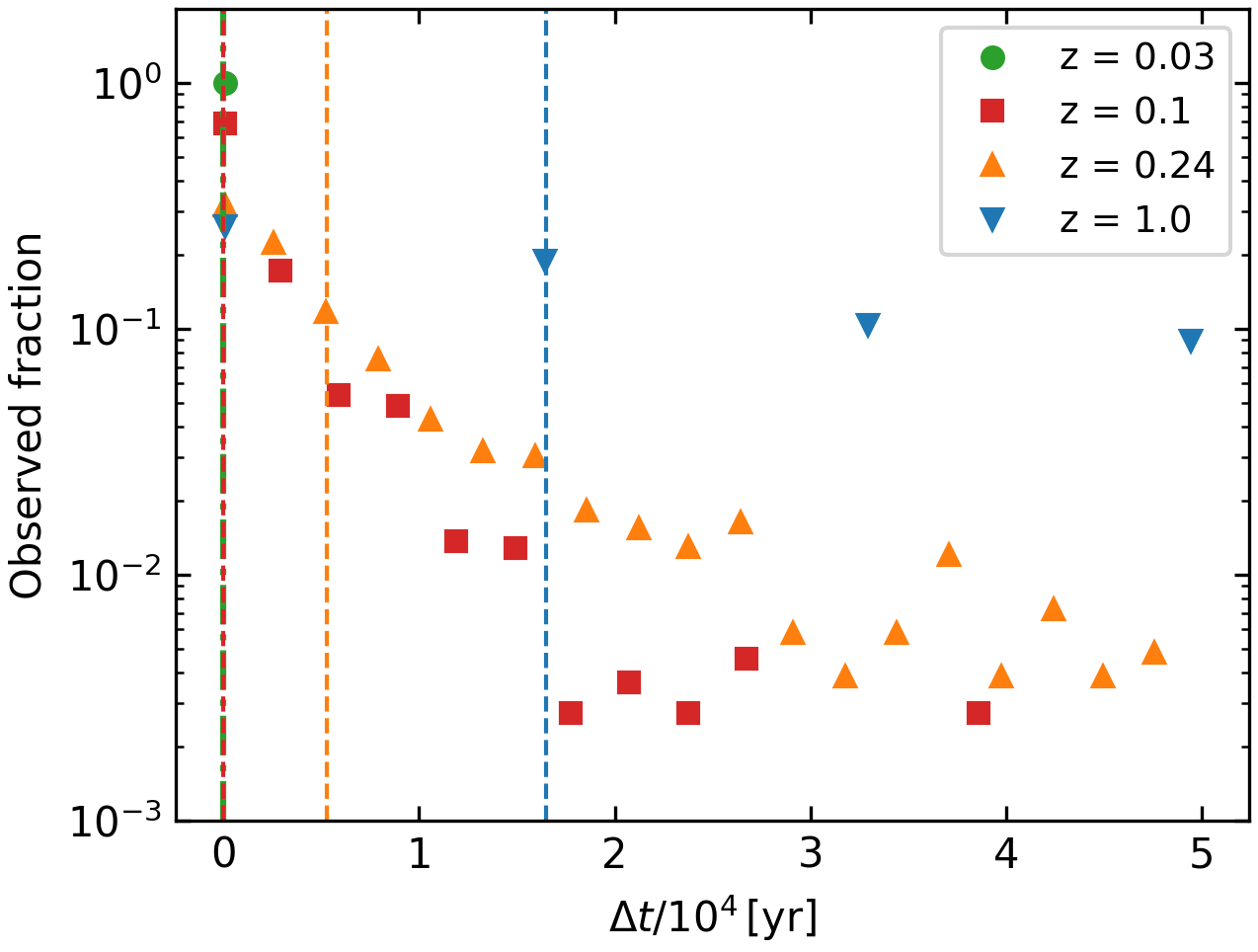}
    \caption{Time delay of neutrinos for different source redshifts. The dashed lines show the arrival times of the fastest $\gamma$-rays from different redshifts, denoted by different colour codes.
    }
    \label{fig:delay_linear}
\end{figure}
\begin{figure*}
    \centering
    \includegraphics[width=0.45\textwidth]{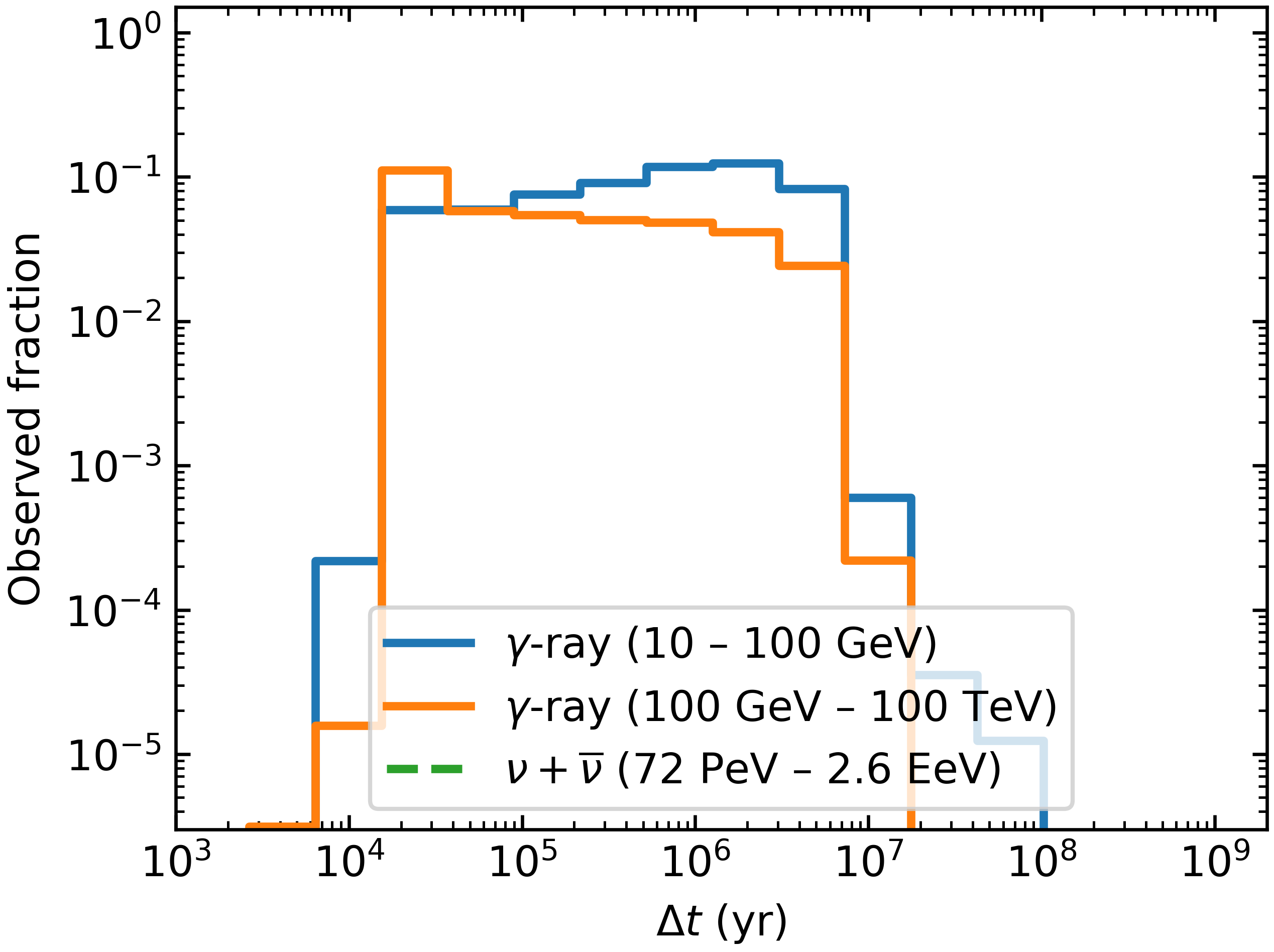}%
    \hspace{1cm}
    \includegraphics[width=0.45\textwidth]{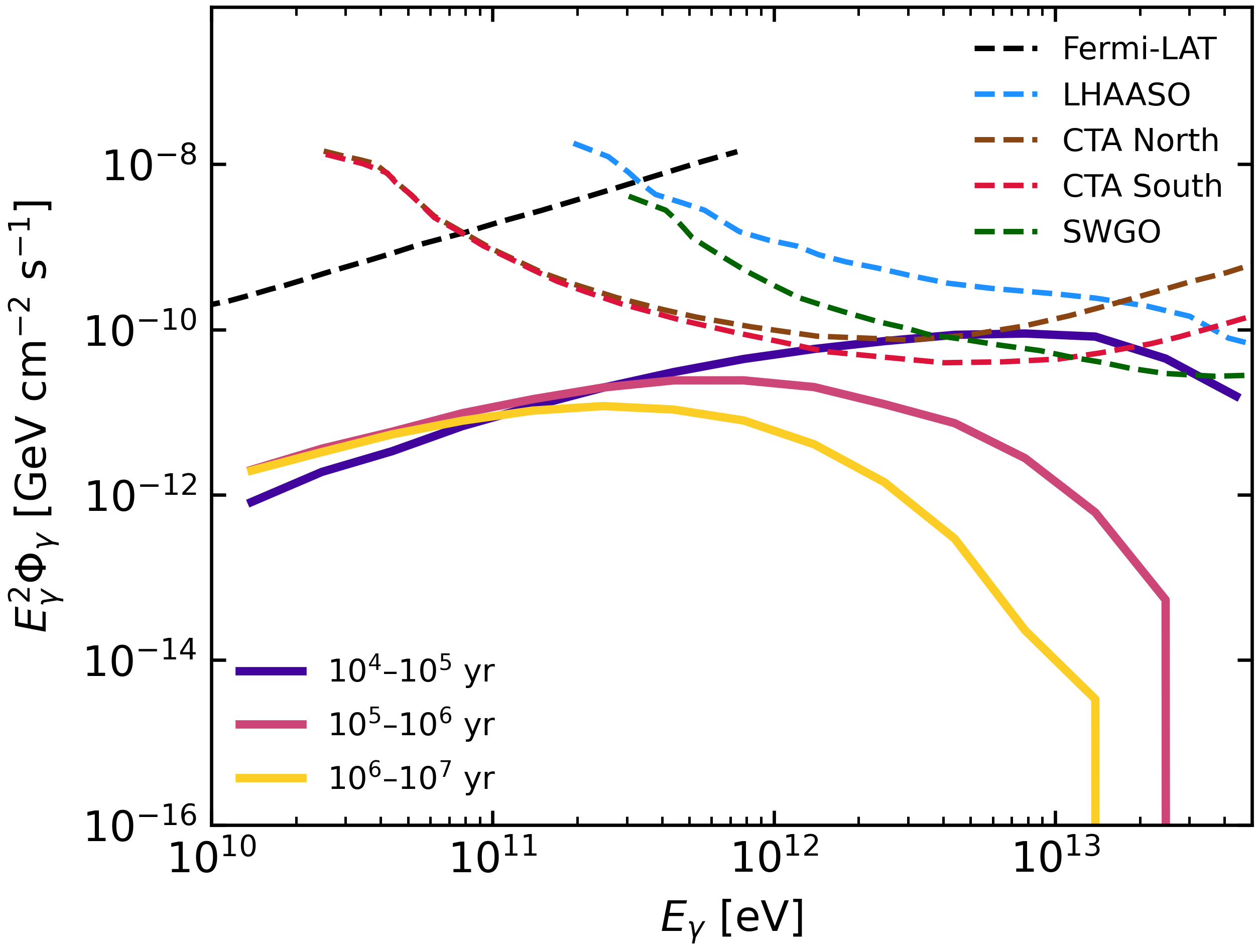}
    \caption{\textit{Left:} Observed neutrino and photon fraction as a function of time delay with respect to the source distance, for a source at $d_c = 133$ Mpc ($z\approx0.01$). The statistics are binned in various energy ranges. For neutrinos, we show the observed fraction corresponding to the central 90\% energy range of the KM3NeT neutrino event. The initial neutrino burst is not included in this case, since all events arrive at $\Delta t = 0$. \textit{Right:} The corresponding $\gamma$-ray spectra for different time delay intervals are shown and compared with the flux sensitivities of $\gamma$-ray telescopes, indicated by dashed curves. The Fermi-LAT sensitivity \citep{bruel2018fermilatimprovedpass8event} corresponds to all-sky performance of the instrument. The CTA North and South \citep{CTAConsortium:2017dvg} assume $50$~hours of exposure, while those for SWGO \citep{Albert:2019afb} and LHAASO \citep{LHAASO:2019qtb} correspond to $5$~year and $1$~year of observation, respectively.}
    \label{fig:d133}
\end{figure*}
\begin{figure*}
\centering
    \includegraphics[width=0.45\textwidth]{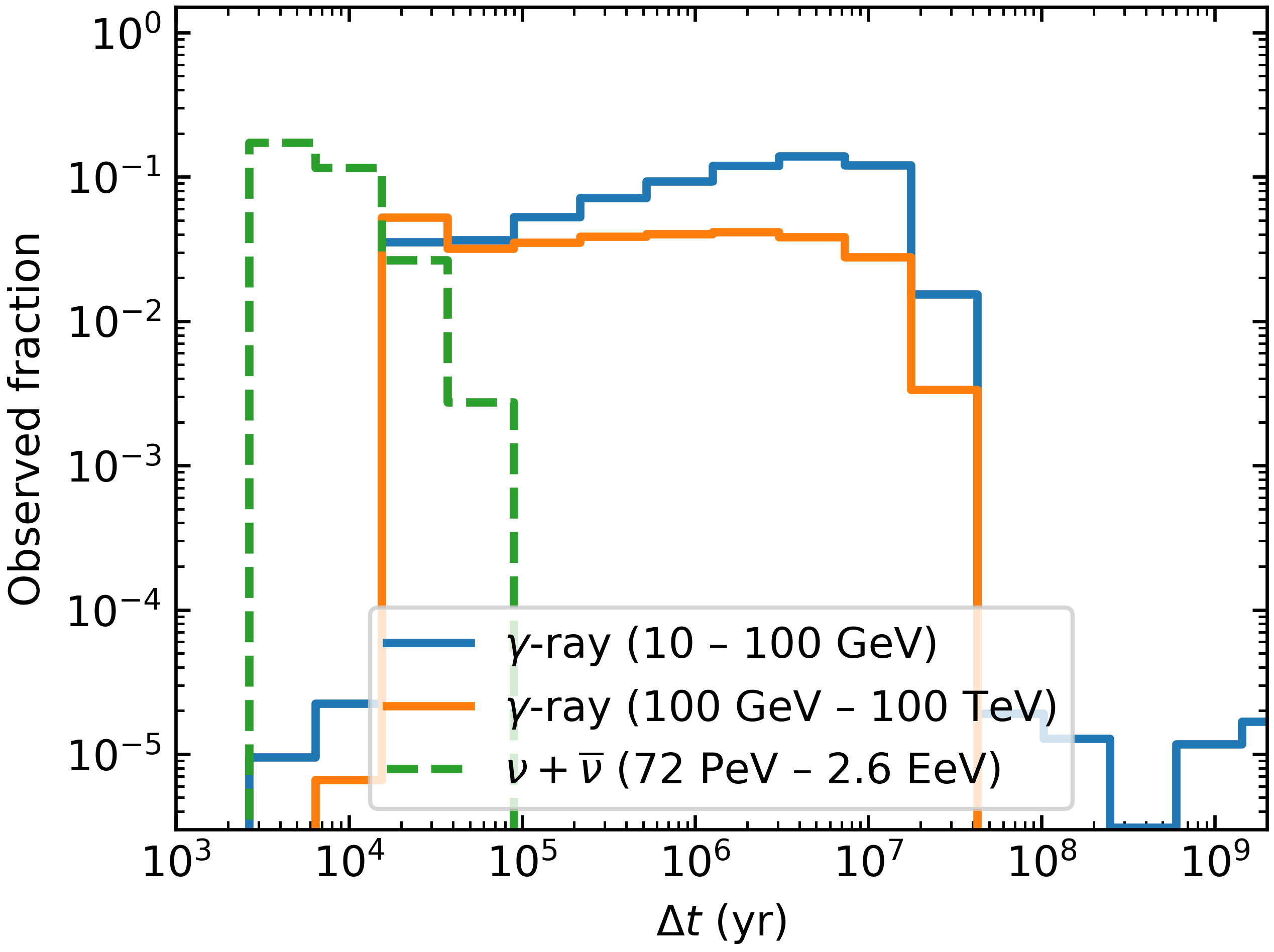}%
    \hspace{1cm}
    \includegraphics[width=0.45\textwidth]{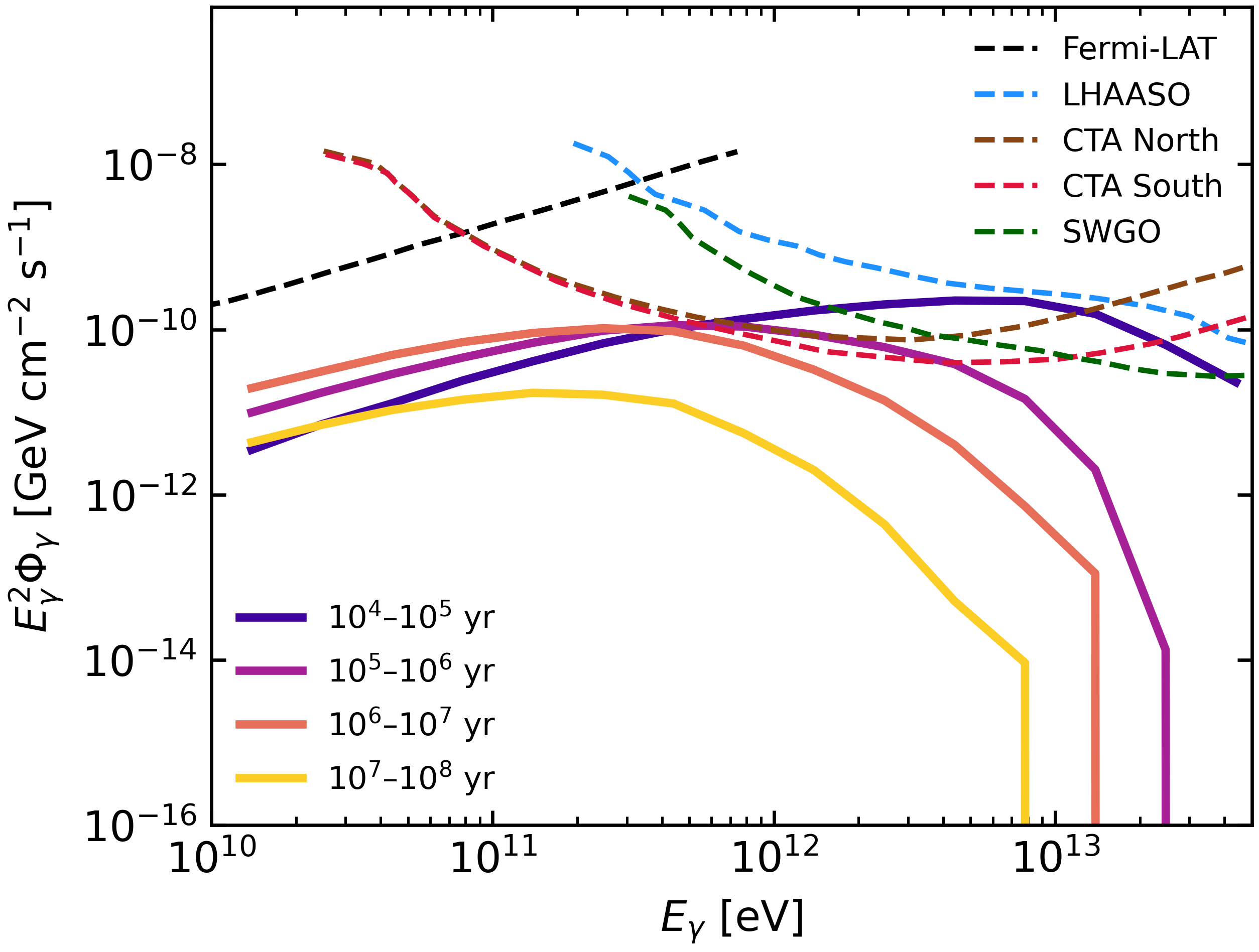}
    \caption{Same as Fig.~\ref{fig:d133} but for comoving source distance $d=435$ Mpc ($z\approx 0.1$) of the transient point source.}
    \label{fig:d435}
\end{figure*}
\begin{figure*}
\centering
    \includegraphics[width=0.45\textwidth]{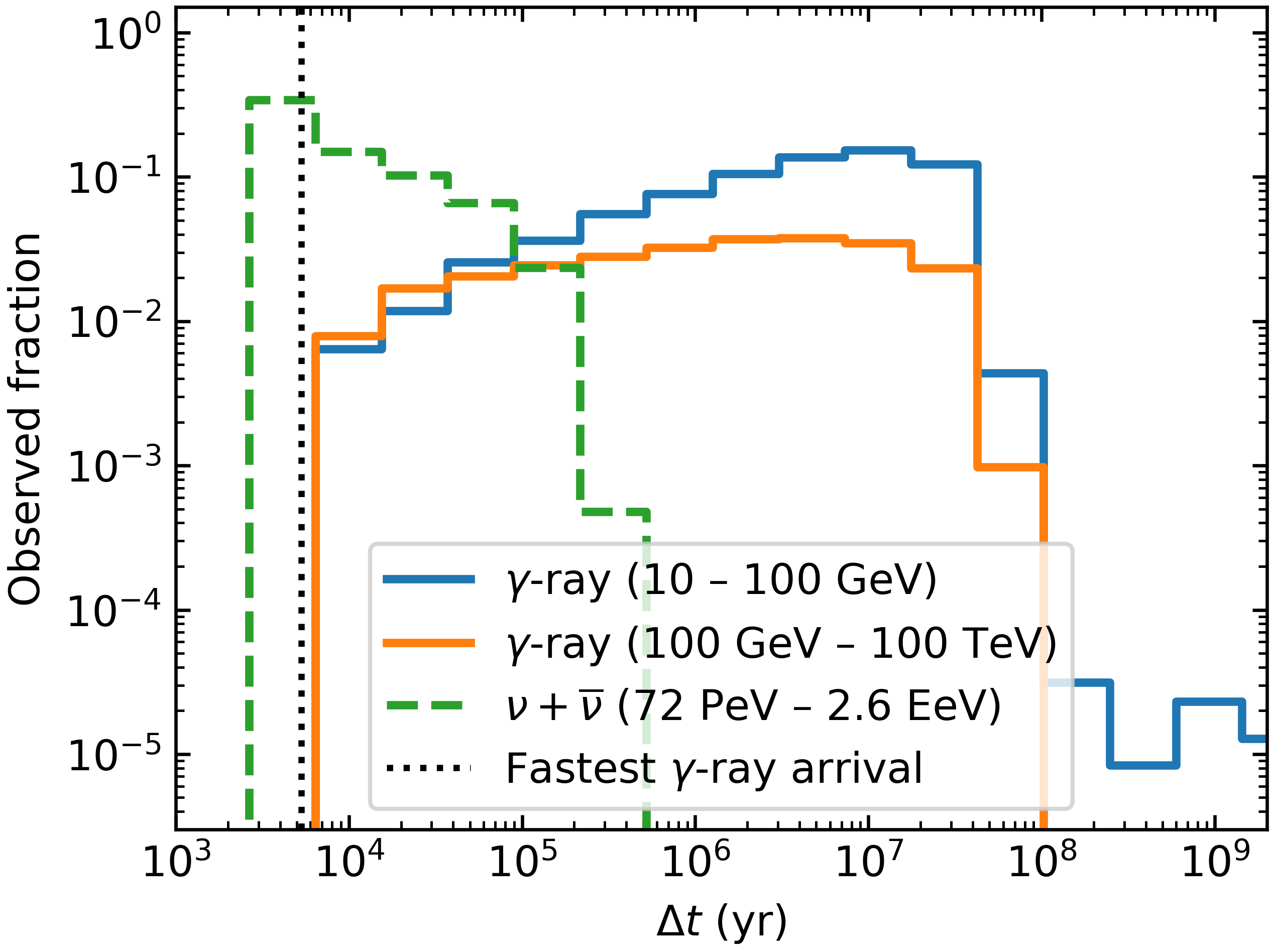}%
    \hspace{1cm}
    \includegraphics[width=0.45\textwidth]{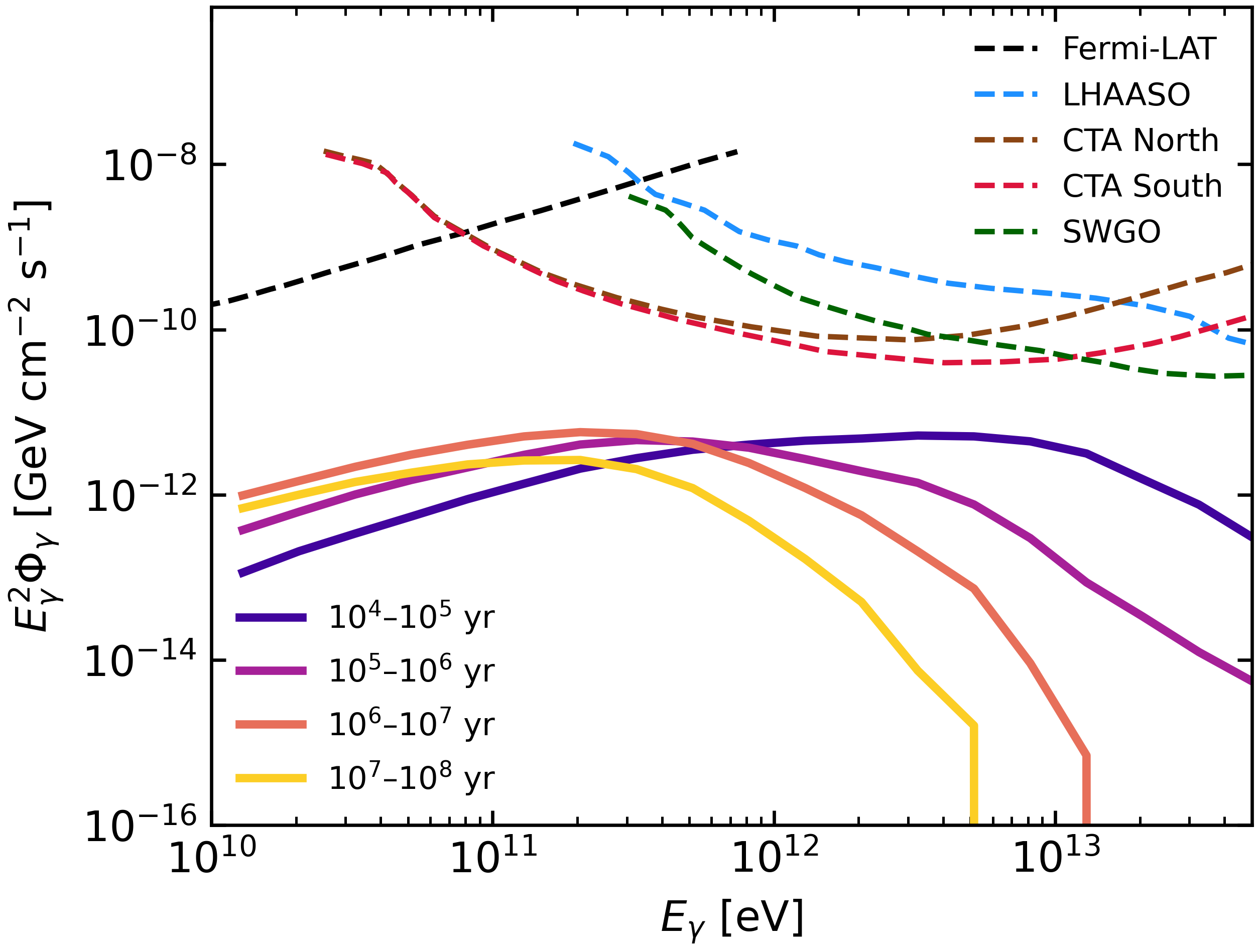}
    \caption{Same as Fig.~\ref{fig:d133} but for comoving source distance $d=1$ Gpc ($z\approx0.24$) of the transient point source.}
    \label{fig:d1000}
\end{figure*}
\begin{figure*}
\centering
    \includegraphics[width=0.45\textwidth]{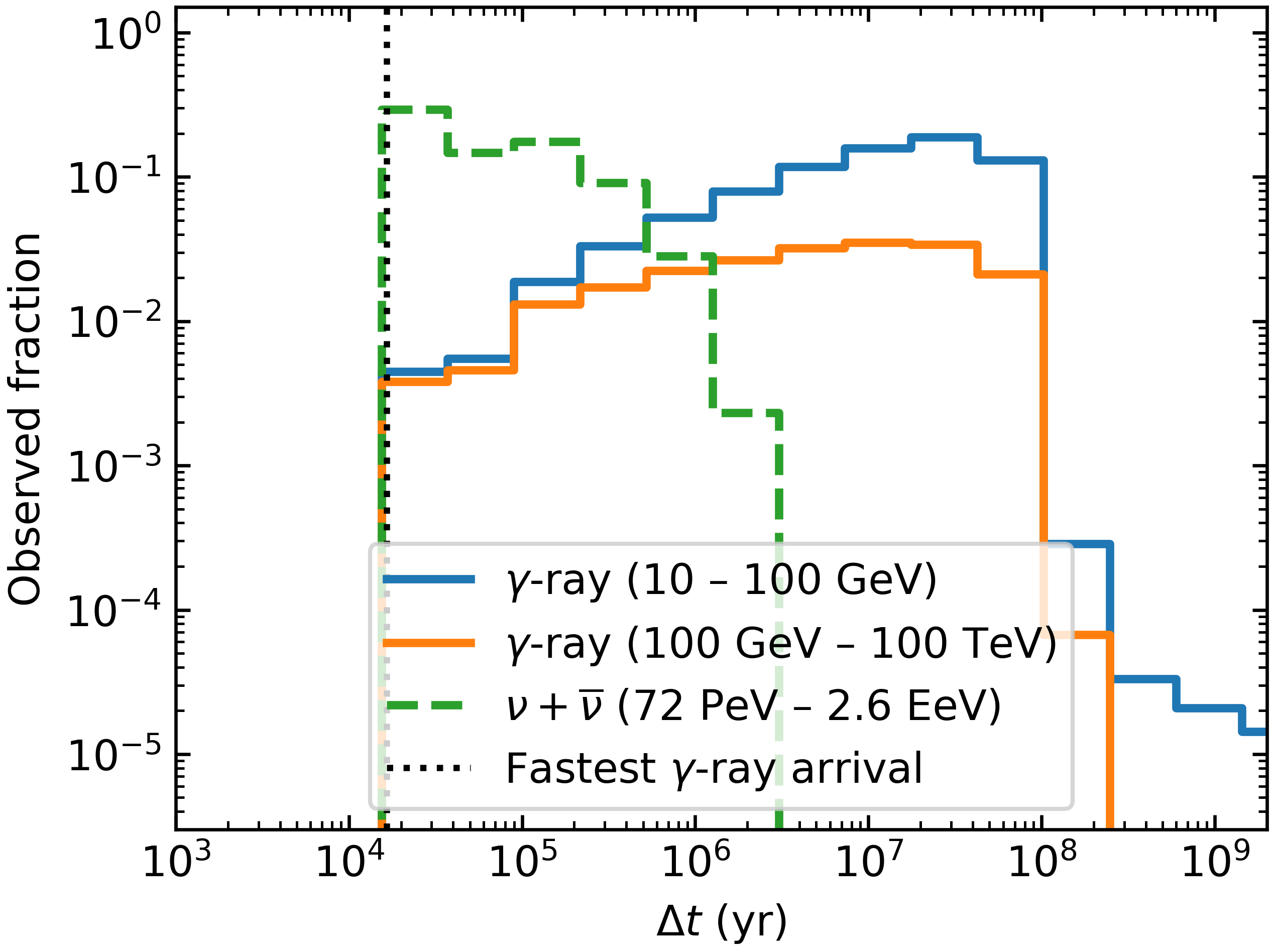}%
    \hspace{1cm}
    \includegraphics[width=0.45\textwidth]{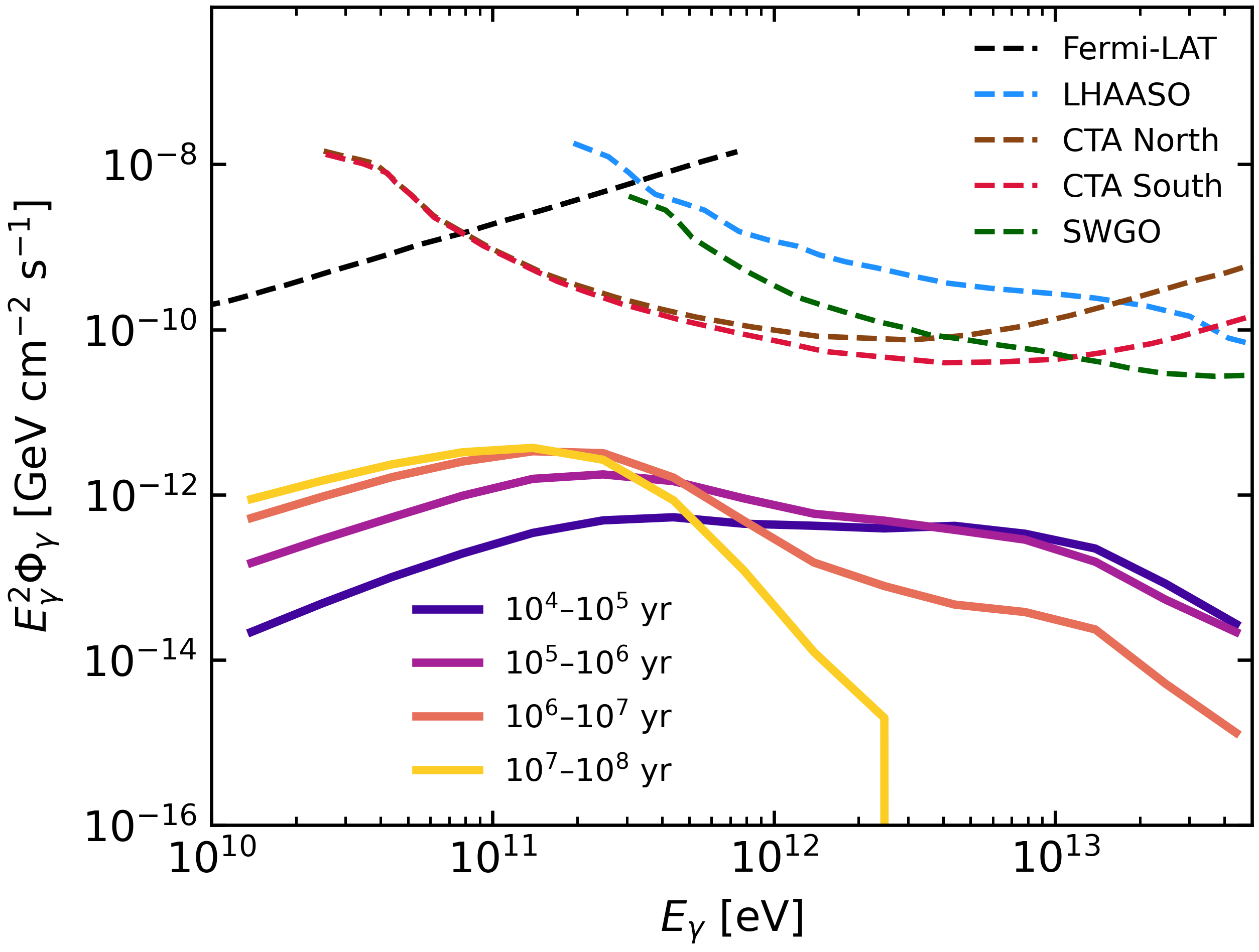}
    \caption{Same as Fig.~\ref{fig:d133} but for comoving source distance $d_c=3.4$ Gpc ($z\approx 1$) of the transient point source.}
    \label{fig:d3400}
\end{figure*}

The distance between the source and the surface of the observer sphere is $d_{\rm eff}$. The time taken by light to travel this distance is needed to calculate the time delay of the neutrinos and $\gamma$-rays produced in cosmic ray interactions. We use the light-crossing time in our time-delay calculations, and source distances are expressed in comoving units unless otherwise noted. Therefore, a delay of $\Delta t = 0$ corresponds to the particle reaching the surface of the observer sphere at light crossing time. For each source redshift, the delay is computed as $\Delta t = (d_{\rm traj } - d_{\rm eff})/c $ where $d_{\rm traj }$ is the total propagation path length of the particle, and assuming that the ultrarelativistic particles travel approximately at the speed of light.
For photons, $d_{\rm traj}$ includes the cumulative path length of all preceding generations in the electromagnetic (EM) cascade from the parent proton. Similarly, for neutrinos, the effective propagation length is the sum of the proton trajectory before the neutrino production and the rectilinear propagation of the neutrino itself. This approach consistently accounts for time delays due to both UHECR propagation and EM cascades when evaluating delays for neutrinos and $\gamma$-rays.
As representative cases over a wide range of distances, we consider source redshifts $z=0.03$, 0.1, 0.24, and 1 for our analysis. At each redshift, we calculated the observed event fraction of $\gamma$-rays and neutrinos, as a function of delay time, denoted by $\Delta t$. We further discriminate the $\gamma$-ray events into energy bins $10-100~\mathrm{GeV}$ and $100~\mathrm{GeV}$--$100~\mathrm{TeV}$, allowing us to analyse the energy and time dependence simultaneously. 

Figure~\ref{fig:delay_linear} presents the time delay distribution of neutrinos (indicated by data points) within the energy range relevant for the KM3NeT event, shown for four different source redshifts.  Since neutinos typically show a significant prompt component, we highlight this behaviour in linear-time binning. For the nearest case $z=0.03$, the entire neutrino population arrives without delay, for our numerical precision. At  $z=0.1$, a substantial fraction reaches the observer at $\Delta t = 0$ while the remaining exhibits noticeable delays with steadily decreasing event fraction at later times. At intermediate reshift $z=0.24$, a prompt component is again visible, followed by a low fraction tail that extends to later times. The qualitative behaviour changes at $z=1$, with the event fraction showing a comparable number of events at delayed times as the prompt arrival, and extending to much later times. This can be attributed to the time delay caused by the deflection of UHE protons in the EGMF.
The earliest $\gamma$-ray arrival times are indicated by the vertical dashed lines, using the same colors as the neutrino delay distribution for individual redshifts, with infinitesimal observed fractions within the time window shown in Figure~\ref{fig:delay_linear}. For low-redshift sources, the earliest photons appear simultaneously with the prompt neutrino population. At higher redshifts, however, a clear temporal separation emerges, due to additional time delay of cosmogenic $\gamma$-rays during the development of the EM cascade. 

The $\gamma$-ray flux peaks at a much later time, as shown in the subsequent Figs.~\ref{fig:d133}-\ref{fig:d3400}. For $z=0.01$, the comoving source distance is $d_c\approx 133$ Mpc and the electromagnetic cascade is not fully developed. As expected for a secondary cascade, the event fraction of lower-energy (10–100 GeV) photons (blue lines) is higher at later times, while at early times the delay distribution (left panels) is comparable to that of higher-energy (100 GeV–100 TeV) photons. This general trend becomes more prominent at higher redshifts, owing to the sufficient cascade of UHE $e^\pm$ and $\gamma$-rays. The right panels show the $\gamma$-ray flux over the entire energy range in our analysis and for various time windows. The fluxes are obtained by normalizing the neutrino spectrum from our simulations to the KM3NeT-only fit flux level at 220 PeV, after accounting for the additional factor due to per-flavor flux after neutrino oscillation. Since a significant fraction of the neutrino events arrive near $\Delta t = 0$, we use the single-event KM3NeT flux for the normalization, with negligible impact on our results. We overlay the sensitivity of currently operating and upcoming telescopes, viz., Fermi-Large Area Telescope (LAT), the northern and southern arrays of Cherenkov Telescope Array Observatory (CTAO) for 50h of observation \citep{CTAConsortium:2017dvg}, Large High-Altitude Air-Shower Observatory (LHAASO) \citep{LHAASO:2019qtb}, and Southern Wide-field Gamma-ray Observatory (SWGO) \citep{Albert:2019afb}. We find that for source distances $z\lesssim 0.2$, the $\gamma$-ray spectrum is comparable to the sensitivities of the upcoming detectors, peaking at $\sim 10$ TeV and time as early as $\Delta t\approx 0.1$ Myr. Whereas, for higher source distances, the cascade is fully developed and high-energy $\gamma$-rays are further absorbed in the EBL, leading to a spectrum that peaks at $\mathcal{O}\sim100$ GeV and at later times. Due to computational limitations, we do not show the $\gamma$-ray fluxes for shorter time intervals.


We propose this trend in the $\gamma$-ray spectrum as an important discriminator of source distance. For past transients that injected UHE protons, the line-of-sight cosmogenic $\gamma$-rays within $1^\circ$ angular deflection and zenith angle within the field of view of imaging atmospheric Cherenkov telescopes should detect orphan $\sim 10$ TeV signals for source distances of a few $\sim$100 Mpc. For higher source distances $z\sim 0.1$, as seen in Figs.~\ref{fig:d435}, the event fractions are comparable in certain $\Delta t$ bins shown for the resolution of our analysis, implying a simultaneous steady flux of neutrinos at sub-EeV and $\gamma$-rays at multi-TeV, maybe correlated. For even higher values of $z$, the neutrino spectrum may evolve with time up to $\Delta t\approx 1$ Myr, as seen in Fig.~\ref{fig:d3400} and also in Fig.~\ref{fig:delay_linear}. This has also been shown in \cite{Zhang:2025abk}. Although the deflection of cosmic rays increases with an increase in source distance, the energy loss length of UHE protons at $\sim10$ EeV is $\sim1$ Gpc and increases for lower energies. This explains the trend obtained in the $\gamma$-ray spectrum and provides crucial insight into the candidate source type of such events.

\begin{figure}
    \centering
    \includegraphics[width=0.45\textwidth]{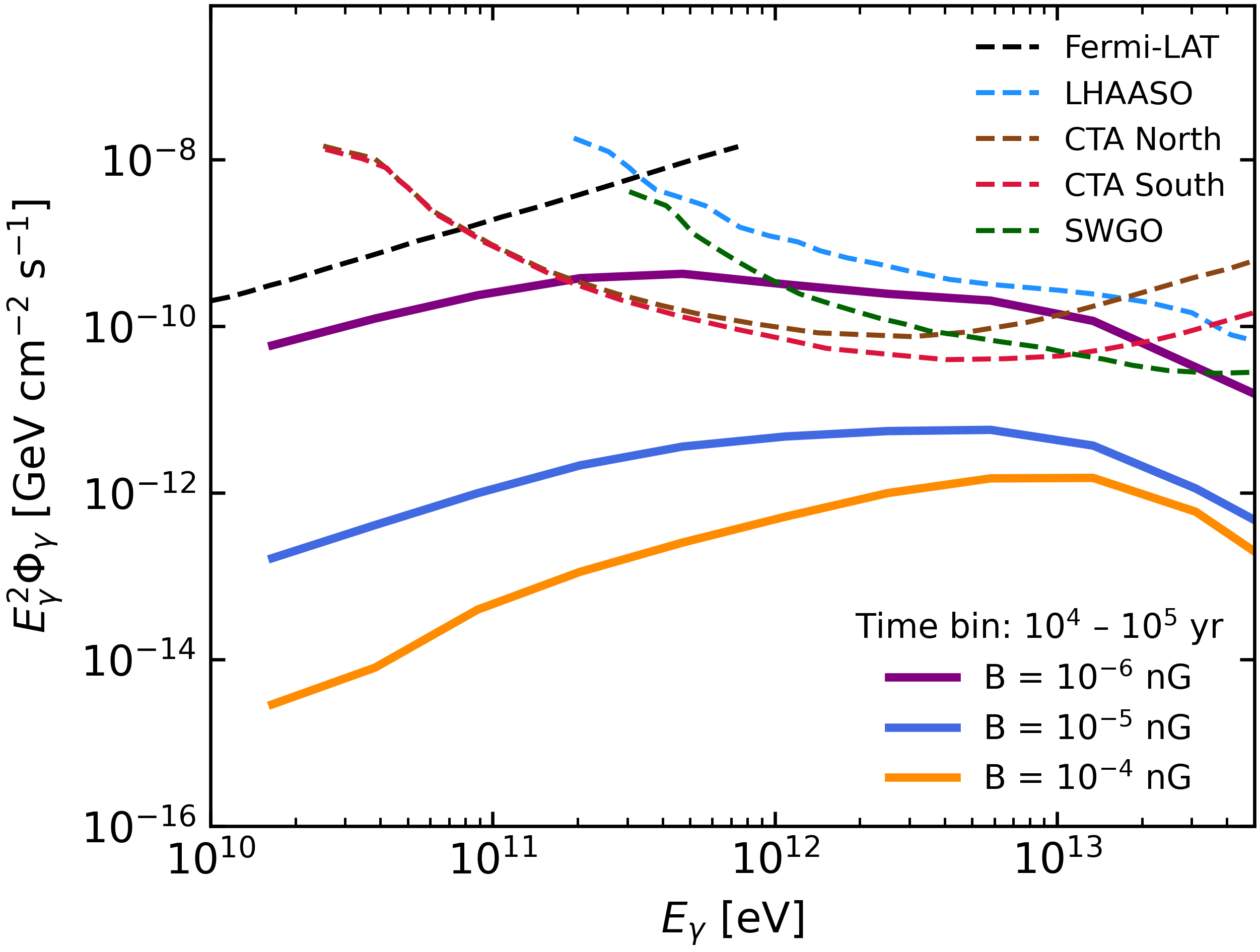}
    \caption{$\gamma$-ray spectra for different EGMF strengths for a source at $z=0.24$ and for a time delay interval between $10^4$ -- $10^5$ years.
    }
    \label{fig:gamma_spectrum_Bfield}
\end{figure}

We also examine how sensitive our results are to variations in the magnetic-field strength. Fig.~\ref{fig:gamma_spectrum_Bfield} shows the $\gamma$-ray flux as a function of the EGMF for a fixed source distance of $d_c = 1$ Gpc (i.e., $z = 0.24$). For a weaker magnetic field of $10^{-6}$ nG, the $\gamma$-ray flux can become comparable to the detector sensitivities at multi-TeV energies. This introduces a degeneracy in the model parameters. However, the detection of multi-TeV $\gamma$-rays would favour a weaker magnetic field or shorter source distance. Since photons originating from inside the source are attenuated by the EBL and cannot reach Earth at such high energies, the detection of such high-energy $\gamma$-rays by future detectors will be a definitive test of the cosmogenic hypothesis. In addition, most prominent cosmic-ray candidate sources exhibit electromagnetic emission. Hence, the present-day detection of such high-energy $\gamma$-rays without an associated multimessenger counterpart would imply the existence of cosmological transients of unknown type in the past that contribute to the diffuse cosmogenic emission, and thus to the Fermi-LAT-measured IGRB.

The sensitivities of current and next-generation facilities are essential for probing such past transients. For cosmological events of this type that occurred $\sim 10^{4}\text{--}10^{6}\,\mathrm{yr}$ ago, the delayed gamma-ray emission may be detectable at the current epoch.
While the accompanying neutrino burst would have passed, the $\gamma$-ray component can be distinguished from the diffuse background due to its hard spectrum at very high energies ($\mathcal{O}\sim10\,\mathrm{TeV}$). Moreover, the peak energy, set by the extragalactic magnetic field, provides additional diagnostic power for source identification. The non-detection of such fluxes by CTAO would imply a transient distance $\gtrsim 500\,\mathrm{Mpc}$. These predictions open a new window to study UHECR production by past cosmological transients.

The analysis presented here, as well as the corresponding spectral shapes, are tailored to an observer sphere that records photons and neutrinos with deflection angles up to $1^\circ$. In view of the angular uncertainty of the KM3NeT event, we therefore focus on fluxes close to the line of sight. By contrast, the diffuse contribution from a population of such transient sources would be dominated by photons with much larger deflections and consequently lower energies. We have also assumed that the initial emission is beamed; however, depending on the source type, the emission could be isotropic. In that case, matching the flux level inferred from the KM3NeT-only fit would require a substantially higher intrinsic cosmic-ray luminosity. We focus on $\gtrsim 1$ TeV $\gamma$-rays that could plausibly be associated with sources inside the sky localisation region. Future detections of sub-EeV neutrinos will provide stronger evidence to discriminate between cosmogenic and astrophysical origins. If the origin is astrophysical, one generally expects an accompanying photon signal in $\gamma$-rays or at other wavelengths.


%

%



\section{Discussions\label{sec:discussions}}

In our work, we explore the cosmogenic $\nu$ and $\gamma$-ray signatures from a transient point source that can account for the flux derived from KM3-230213A detection. We assume the source accelerates UHE protons up to $\approx10$ EeV. The dominant energy-loss process for protons below $E\lesssim 2\times10^{18}$ eV is that due to adiabatic expansion of the Universe. The combined fit of the Auger spectrum and composition suggests a light nuclei composition at $\sim 1$ EeV with progressively heavier composition at $\gtrsim 10^{18.2}$ eV \citep{PierreAuger:2016use}. Beyond the ankle at $\gtrsim 10^{18.7}$ eV, the sources must originate from redshifts $z\lesssim 0.5$ to survive photodisintegration of heavier nuclei. An additional source population with pure proton injection and spectral index $\gtrsim2$ has been proposed to account for the diffuse UHECR spectrum \citep{Muzio:2019leu, Das:2020nvx}. This justifies our choice of the mass and energy spectrum injected by the transient. We assume it belongs to such a distinct subdominant source population, so that the cumulative contribution remains compatible with the diffuse IGRB background.

Our numerical analysis incorporating 3-D deflections of UHE protons in the EGMF reveals a time delay of cosmogenic $\gamma$-rays that peaks at around $\sim 100$ kyr with respect to the light crossing time. This is consistent with the analytical estimates of the time delay due to inverse-Compton cooling and bending of the electron trajectory in the EGMF. Hence, if the origin of KM3-230213A is truly cosmogenic, the time evolution of the $\gamma$-ray spectrum shows that a detection at the current epoch is challenging. However, for similar cosmological transients in the past, the gamma-ray spectrum peaks at $\sim 10$ TeV for a source at $z\lesssim 0.2$. For a source at $\sim0.1$, as seen in Fig.~\ref{fig:d435}, the event fraction of 100 GeV -- 100 TeV photons is significant at $\Delta t \sim 10^4$, when the neutrino event rate declines. At higher source distances, the neutrino spectrum evolves over Myr time-delay scales within the 90\% central energy range of the KM3NeT event. However, at higher redshifts, the peak energy of the $\gamma$-rays shifts to lower energy, as expected for a fully developed EM cascade, and is below the detection threshold for a prudent choice of EGMF. This implies multi-TeV $\gamma$-ray and sub-EeV neutrino flux may be correlated for comoving source distances $\lesssim 500$ Mpc. The effects of the Galactic magnetic field can be neglected here, since nearly all secondaries are produced at distances from Earth much larger than the virial radius of our Galaxy and well beyond the observer size in our simulations.

For known transient sources such as radio-loud AGNs, a significant electromagnetic counterpart is expected. Earlier studies of blazars invoking cosmogenic $\gamma$-rays indicate appreciable interactions within the sources \citep[see, e.g.,][]{Keivani:2018rnh, MAGIC:2022gyf, Das:2022nyp} leading to a considerable hadronic component in X-rays and $\gamma$-rays. Similarly, in the case of gamma-ray bursts (GRBs), a prompt $\gamma$-ray/X-ray emission or extended multiwavelength afterglow emission is expected. The non-detection of high-energy neutrinos from GRBs by IceCube \citep{IceCube:2014jkq, IceCube:2017amx} has severely constrained cosmic-ray acceleration during the prompt emission phase. Sub-threshold $\gamma$-ray sources, including transients, that are not included in Fermi catalogues have been proposed \citep{Sherman:2025gir}. For radio-quiet AGNs or Seyfert galaxies, reprocessed GeV-TeV $\gamma$-rays contribute to a significant X-ray emission \citep{Kheirandish:2021wkm, Murase:2022dog}. Hence, the time delay between a prompt neutrino produced near the source and its electromagnetic counterpart may pose difficulties with known transient models \citep{Das:2025vqd}. Thus, the exact nature of the transient is difficult to derive without an electromagnetic counterpart. However, the proximity of the KM3-230213A event to the Galactic plane may have hindered source identification due to Galactic foreground emission.


\section{Conclusions\label{sec:conclusion}}

The recent detection of KM3-230213A provides an interesting case study of the predicted neutrino flux level from a transient point source due to UHECR interaction with the cosmic infrared background. Similar transient phenomena in the past will produce delayed cosmogenic $\gamma$-ray flux. We present a time-dependent analysis of the observed signal, using 3-D numerical simulations, to identify the discriminatory signatures of source distance and the nature of the transient, given that multi-TeV photons will saturate the upcoming generation of $\gamma$-ray detectors. The time-delay distribution of the electromagnetic cascade is directly related to the extragalactic magnetic field strength, and the detection of a steady flux will constrain it using the peak energy. Our results suggest that the detection of multi-TeV $\gamma$-rays would imply a transient source with a low $p\gamma$ optical depth, suppressing interactions within the source. The lack of electromagnetic counterparts at other wavelengths makes it challenging to explain with known transients. We conclude that for a comoving distance of a few hundred Mpc, the multi-TeV $\gamma$-ray flux may be correlated with the sub-EeV cosmogenic neutrino spectrum.

\section{Acknowledgement}
S.B. and N.G. thank R. Alves Batista for helpful discussions on estimating trajectory lengths with CRPropa3. We thank the anonymous referee for valuable comments to improve the manuscript.
\software{CRPropa3 (\url{http://ascl.net/2208.016})}

\bibliography{reference}{}
\bibliographystyle{aasjournal}
\end{document}